\documentclass[11pt]{article}
\usepackage{aas2pp4}
\slugcomment{Accepted for publication in the Astrophysical Journal}
\begin{document}
\title{EGRET Gamma-Ray Observations of the Crab P2/P1 Ratio}
\author{W. F. Tompkins\altaffilmark{1}, B.B. Jones, P.L. Nolan}
\affil{W. W. Hansen Experimental Physics Laboratory, Stanford University, Stanford CA 94305}
\authoraddr{Stanford University, Stanford CA 94305}
\author{G. Kanbach}
\affil{Max Planck Institut f\"ur Extraterrestische Physik, D85748 Garching, Germany}
\authoraddr{D85748 Garching, Germany}
\and
\author{P. V. Ramanamurthy, D. J. Thompson}
\affil{NASA/Goddard Space Flight Center, Code 661, Greenbelt MD 20771}
\authoraddr{Code 661, Greenbelt MD 20771}
\altaffiltext{1}{billt@egret0.stanford.edu}
\date{}
%\maketitle

\begin{abstract}
Recent  observations of the Crab pulsar by the Energetic Gamma-Ray Experiment
Telescope (EGRET) on the Compton Gamma Ray Observatory show that the high-energy gamma-ray light curve
has changed little over the lifetime of the instrument.  Previous data collected by SAS-2 and COS-B
in the years 1972-82, along with earlier EGRET data, suggested a 14 year
sinusoidal variation  in the flux ratio between the first and second peaks.
The new data from EGRET 
indicate that the flux ratio is constant.  
\end{abstract}


\begin{center}
Accepted Astrophysical Journal.  Scheduled 20 Sep 1997.
\end{center}

\twocolumn

\section{Introduction}
High energy gamma ray emission from the Crab pulsar was observed by satellite-borne
telescopes for 15 years: in 1972-73 by SAS-2 (Kniffen et al. 1974),
from 1975-82 by COS-B (Clear et al. 1987),
and since 1991 by EGRET (Nolan et al. 1993; Ramanamurthy et al. 1995).
Early observations showed possible sinusoidal  variation in the relative
intensities of the
two peaks (Wills et al. 1982) with a time scale of $\sim 14$ years,
and it was suggested that this variation might
be due to the precession or free nutation of the neutron star
(Kanbach 1990, \"Ozel 1991).  An apparent confirmation of the sinusoidal
signal was seen in the low energy
gamma ray emission (Ulmer et al. 1994), matched
in phase and period with the high energy results, but with a smaller amplitude.
EGRET data from 1991 through early 1994 were consistent with the expected
variation (Nolan et al. 1993, Ramanamurthy et al. 1995), although these observations
spanned a time when the ratio of the peaks was predicted to be fairly constant, near
the minimum of the sinusoid.

EGRET observations have now extended the available data by over two years.  The
most recent data were expected to be $4 - 6 \sigma$ from the
average of the previous values if the sinusoidal model is correct.

\section{Observations and Analysis}

The EGRET instrument is a spark chamber gamma-ray telescope with an energy range of
30~MeV -- 30~GeV.  Details of the instrument design, calibration, and 
standard analysis software are given in Thompson et al. (1993).

All viewing periods where EGRET was pointed within $20 ^{\circ}$ of the Crab
were analyzed, with the exception of viewing period 0021 (1991 Jul 8--15),
in which there was a large solar flare.  The Compton Observatory viewing period
numbers and dates for these observations are shown in Table 1.  The eight viewings
numbered 4120 through 5280 (1995 Feb. through 1996 Aug.) have been completed since
the time of the previous work of Ramanamurthy et al. (1995). 

All photons with measured energy
above 50 MeV, which were within an energy-dependent cone of half-angle
$\theta_{max}$ were used in this analysis.  The angle $\theta_{max}$,
chosen such that $68\%$ of the photons originating from the pulsar are within
the acceptance cone, is given by (Thompson et al. 1993)
\begin{displaymath}
\theta_{max} = 5^{\circ}.85 \times (E/100 MeV)^{-0.534}.
\end{displaymath}
\noindent The arrival time of each detected photon was transformed to Solar System
Barycentric Time using the DE200 ephemeris, then binned according to the pulsar
phase at that time,
determined from the Princeton Pulsar Timing Database (Arzoumanian et al. 1992).
This analysis was performed using the PULSAR program (Fierro 1995).

As seen in Figure 1, the light curve was divided into several sections,
including Peak 1 (phase .94 -- .04), Peak 2
(phase .32 -- .46), and the off-pulse Background (phase .46 -- .94).
These
definitions follow those used in the COS-B analysis (Wills et al. 1982) and are
similar to those used by Nolan et al. (1993) and Ramanamurthy et al. (1995).  The
background (from the Crab nebula, nearby sources, and the diffuse Galactic radiation)
was assumed constant as a function of pulsar phase.  The off-pulse count rate was
then used to find the background-subtracted
counts estimates of the two peaks ($P1$ and $P2$).
In order to avoid any effects of changes in instrument performance, the evolution
of the ratio $P2/P1$ was examined (as in previous analyses).  

\begin{figure}[htb]
\plotone{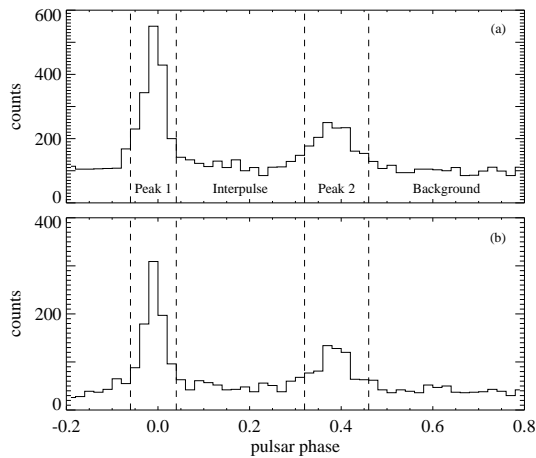}
\caption{\footnotesize Gamma-ray phase histograms of photons with $E>50$ MeV for EGRET observations of the Crab in 1991-2 ({\it a}) and 1995-6 ({\it b}).
The vertical dashed lines
indicate phase boundaries used in the peak height analysis.  The horizontal
dashed lines indicate the background level as determined from the data.}
\end{figure}

The two peaks have slightly different energy spectra (Nolan et al. 1993), and
the EGRET response at different energies has changed at different rates
(Esposito et al. 1997).
Thus the ratio $P2/P1$ is affected by the changes in instrument performance
over time.  Calculations of this effect, however, indicate that it is an order
of magnitude smaller than the errors in $P2/P1$ due to Poisson fluctuations.

The differences in the energy responses of the SAS-2, COS-B, and EGRET
instruments are larger than the variation in the EGRET response.
However, statistical errors in the previous
instruments' data are larger as well. Thus the value of $P2/P1$ obtained from
EGRET data should be comparable to that obtained
with SAS-2 and COS-B.

\section{Results}

The values of $P2/P1$ obtained, together with the $1 \sigma$ errors,
are shown in Table 1.
A reduced data set,
where nearby points are joined for clarity, is shown in Figure 2.
The data were fit
with a constant, yielding $P2/P1 = .54 \pm .03$, with $ \chi^2 = 6.01 $
with 20 degrees of freedom (DOF).  Such a low value of $\chi^2$ might imply
that the errors in the data were over-estimated.  In this case, however,
the errors arise purely from statistical Poisson fluctuations, and the
low value must occur purely by chance.  The
data are very consistent with a constant value of $P2/P1$.
\begin{figure}[htb]
\plotone{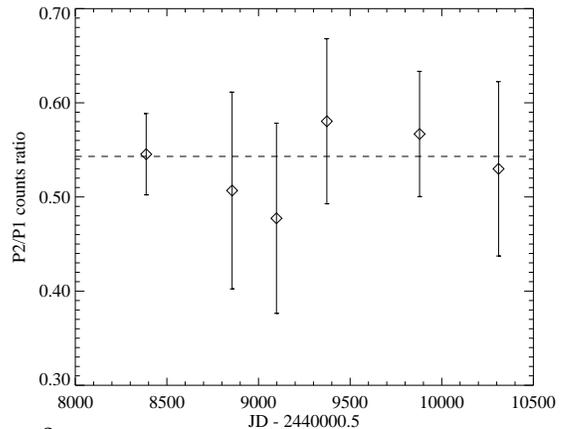}
\caption{\footnotesize Variation in the ratio of the two peaks in the Crab light curve
for $E>50$ MeV (from Table 1).  For clarity, the 21 observations are grouped
into 6 data points, where each point represents the average of several nearby
observations.  Error bars are $1 \sigma$.
The dashed line is the average of all EGRET observations.}
\end{figure}
The data were also fit with a straight line ($\chi^2 = 5.96$ with 19 DOF),
and a quadratic ($\chi^2 = 5.92$ with 18 DOF).  Neither result gives 
a significantly better fit: the EGRET data are most consistent
with no variation.

The EGRET data were also analyzed in conjunction
 with the SAS-2 (Kanbach 1990) and
COS-B (Clear et al. 1987) data.  The best fit
 sinusoid to the previous instruments' data,
\begin{displaymath}
P2/P1 = 0.85 - 0.56 \sin \left(2 \pi \left(T - 1975.67\right)/13.3\right),
\end{displaymath}
where T is the year of the observation, 
and to the combined data set,
\begin{displaymath}
P2/P1 = 0.544 - 0.060 \sin \left(2 \pi \left(T - 1976.48\right)/11.55\right)
\end{displaymath}
are shown together with the data and the average value of P2/P1 in Figure 3.

\begin{figure}[htb]
\plotone{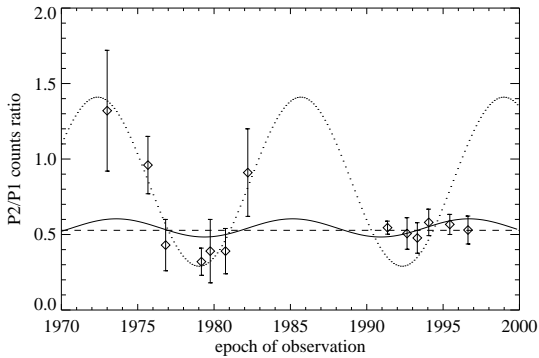}
\caption{\footnotesize Variation in the ratio of the two peaks in the Crab light curve
from SAS-2 (1973), COS-B (1975-1983), and EGRET (1991-96), where the EGRET
data set has been reduced as in Fig 2.  The dotted line
is the best fit sinusoid to the pre-EGRET data.  The solid line is the best
fit sinusoid to all the data.  The dashed line is the average of all
the data.}
\end{figure}

As can be seen, the most recent EGRET observations (the last data point)
are much less consistent with the
large amplitude sinusoid.  The constant value of $0.528 \pm 0.027$ gives a
$\chi^2 = 24.0 $ (with 27 DOF), indicating a good fit.  The sinusoid fit to the
combined data gives a period of 11.6 years with 
$\chi^2 =  21.2$ (with 24 DOF), which does not represent a significant
improvement.  Thus, using the
combined data sets, the data are most consistent with no variation in P2/P1.

The light curves obtained from Phase 1 data (Apr 91--Sept 92)
and from Phases 4 and 5 (Feb 95--Aug 96) are shown in Figure 1.  The
overall shape seems to have changed little, in contrast with the expected
change if the 14 year cycle were correct.  A $\chi^2$ test was performed
to compare the two light curves in a quantitative way.
In order to take out possible systematic effects, both an additive offset
and a multiplicative
factor for the second light curve were fit to the data.  The resulting
$\chi^2 = 58.5$ (with 48 DOF) is consistent with no change in the light curve.
Without the offset, $\chi^2 = 87.9$ (with 49 DOF) was obtained, indicating
an inconsistency at the 99.95\% confidence level.  The offset required
indicates a lower background level in the later observations.  This could be due
to a change in the nebular emission (de Jager et al. 1996), or could be a
result of changes in the performance of EGRET.  As the gas in the spark chamber
ages, the sensitive area at low energies decreases,
which decreases the width of the
average point spread function.  This effect might lower the background
in the later observations.

\section{Summary and Conclusion}
Recent observations with EGRET have provided data relevant to the reported
variation in the P2/P1 ratio of the Crab pulsar.  Data from SAS-2 and COS-B
suggested a sinusoidal variation in this ratio.  The EGRET data, both taken
alone and in conjunction with the data from previous instruments, are most
consistent with a constant value of P2/P1.  Examination of the light
curves
from early and later observations shows no distinct changes in the pulsar's
light curve.  The EGRET data cannot, of course,
rule out past variability in the P2/P1 ratio.  Future observations by EGRET
and its successors may allow a more precise characterization of the long
term behavior of the Crab's light curve.

The EGRET team gratefully acknowledges support from the following:
The ARCS Foundation (WFT),
Bundesministerium f\"ur Bildung, Wissenschaft,
Forschung und Technologie grant 50 QV 9095 (MPE),
NASA Cooperative Agreement NCC 5-95 (HSC),
NASA Grant NAG5-1605(SU),
and NASA Contract NAS5-96051 (NGC).

\onecolumn
\begin{deluxetable}{lrl}
\tablecaption{Crab P2/P1 for Each EGRET Observation \label{mybigtable}}
\tablehead{
\colhead{Viewing Period} &
\colhead{Dates} &
\colhead{P2/P1} 
}
\startdata 
0002 & Apr 22--28 1991 & $  0.57  \pm    0.11$ \nl
0003 & Apr 28--May 1 1991 & $  0.54  \pm    0.15$ \nl
0004 & May 1--4 1991 & $  0.65  \pm    0.18$ \nl
0005 & May 4--7 1991 & $  0.55  \pm    0.14$ \nl
0010 & May 16--30 1991 & $  0.52  \pm   0.07$ \nl
0360 & Aug 11--12 1992 & $  0.66  \pm    0.34$ \nl
0365 & Aug 12--20 1992 & $  0.37  \pm    0.15$ \nl
0390 & Sep 1--17 1992 & $  0.65  \pm    0.17$ \nl
2130 & May 23--29 1993 & $  0.45  \pm    0.19$ \nl
2210 & May 13--24 1993 & $  0.49  \pm    0.12$ \nl
3100 & Dec 1--13 1993 & $  0.55  \pm    0.16$ \nl
3211 & Feb 8--15 1994 & $  0.59  \pm    0.11$ \nl
3215 & Feb 15--17 1994 & $  0.61   \pm   0.31$ \nl
4120 & Feb 28--Mar 7 1995 & $  0.52   \pm   0.15$ \nl
4130 & Mar 7--21 1995 & $  0.58   \pm   0.12$ \nl
4200 & May 23--Jun 6 1995 & $  0.45   \pm   0.15$ \nl
4260 & Aug 8--22 1995 & $  0.50   \pm   0.20$ \nl
5020 & Oct 17--31 1995 & $   0.73  \pm    0.14$ \nl
5260 & Jul 30--Aug 13 1996 & $   0.50  \pm    0.12$ \nl
5270 & Aug 13--20 1996 & $   0.69   \pm   0.21$ \nl
5280 & Aug 20--27 1996 & $   0.48  \pm    0.22$ \nl
\enddata
\end{deluxetable}


\begin{references}

\reference{}Arzoumanian. Z., Nice, D., \& Taylor, J.H. 1992, GRO/radio timing data base, Princeton University
\reference{}Clear, J. et al. 1987, \aap, 174, 85
\reference{}Esposito, J.A. et al. 1997, in preparation.
\reference{}de Jager, O.C. et al. 1996, \apj, 457, 253
\reference{}Fierro, J. M. 1995, Ph.D. Thesis, Stanford University
\reference{}Kanbach, G. 1990, in The EGRET Scinece Symposium: NASA Conference
Publication No. 3071 ed. Fichtel, C.E. et al. 101
\reference{}Kniffen, D.A. et al. 1974 \nat, 251 397
\reference{}Nolan, P.L. et al. 1993, \apj, 409, 697
\reference{}\"Ozel, M.E. 1991, Europhysics Letters, 14, 3
\reference{}Ramanamurthy, P.V. et al, 1995, \apj, 450, 791
\reference{}Thompson, D.J. et al. 1993, \apjs, 86, 629
\reference{}Ulmer, M.P. et al. 1994, \apj, 432, 228
\reference{}Wills, R.D. et al. 1982, Nature, 296, 723

\end{references}
\end{document}